\documentclass[twocolumn,superscriptaddress,longbibliography,aps,pra,preprintnumbers]{revtex4-1}

\usepackage{graphicx}
\usepackage{bm}
\usepackage{color}
\usepackage{epstopdf}
\usepackage{amsmath}
\usepackage{amssymb}
\usepackage{epstopdf}
\usepackage[urlcolor=blue,colorlinks=true,citecolor=blue,linkcolor=blue,pdfstartview={FitH},bookmarks=false]{hyperref}

\graphicspath{{fig/}{./fig/}{.}}

\begin{document}

\title{Interplay between pairing and correlations in 
spin-polarized bound states}

\author{S. G\l{}odzik}
\affiliation{Institute of Physics, M.\ Curie-Sk\l{}odowska University, 
20-031 Lublin, Poland}

\author{A. Kobia\l{}ka}
\affiliation{Institute of Physics, M.\ Curie-Sk\l{}odowska University, 
20-031 Lublin, Poland}

\author{A.\ Gorczyca-Goraj}
\affiliation{Institute of Physics, University of Silesia, 41-500 Chorz\'ow, Poland}

\author{A. Ptok}
\affiliation{Institute of Nuclear Physics, Polish Academy of Sciences, 
31-342 Krak\'{o}w, Poland}

\author{G. G\'orski}
\affiliation{Faculty of Mathematics and Natural Sciences, University of Rzesz\'ow, 
       35-310 Rzesz\'ow, Poland}

\author{M. M.\ Ma\'ska}
\affiliation{Institute of Physics, University of Silesia, 41-500 Chorz\'ow, Poland}

\author{T. Doma\'{n}ski}
\email[e-mail: ]{doman@kft.umcs.lublin.pl}
\affiliation{Institute of Physics, M.\ Curie-Sk\l{}odowska University, 
20-031 Lublin, Poland}

\date{\today}

\begin{abstract}
We investigate the single and multiple defects embedded in a superconducting 
host, studying interplay between the proximity induced pairing and interactions. 
We explore influence of the  spin-orbit coupling on energies, polarization and 
spatial patterns of the bound (Yu-Shiba-Rusinov) states of magnetic impurities 
in 2-dimensional square lattice. We also address the peculiar bound states in 
the proximitized Rashba chain, resembling the Majorana quasiparticles, focusing 
on their magnetic polarization which has been recently reported by [S. Jeon {\it et al.}, 
\href{http://science.sciencemag.org/content/early/2017/10/11/science.aan3670}
{Science {\bf 358}, 772 (2017)}]. Finally, we study leakage of these polarized 
Majorana quasiparticles on the side-attached nanoscopic regions and confront
them with the subgap Kondo effect near to the singlet-doublet phase transition.
\end{abstract}

\maketitle

\section{Introduction}
\label{sec.intro}

Magnetism is usually detrimental to superconductivity because 
it breaks the Cooper pairs (at critical $H_{c2}$). There are, however,  
a few exceptions when these phenomena coexist e.g.\ in iron pnictides 
\cite{Choi-2017}, CeCoIn$_{5}$ \cite{Kenzelmann-2008} or sometimes magnetic 
field induces superconductivity \cite{Meul-1984}. Plenty of other interesting 
examples can be found in nanoscopic systems, where magnetic impurities (dots) 
have more subtle relationship with the electron pairing driven by the proximity effect
\cite{balatsky.vekhter.06,heinrich.pascual.17}.  Cooper pairs 
easily penetrate the nanoscopic impurities, inducing the bound 
(Yu--Shiba--Rusinov) states that manifest the local pairing coexisting
with  magnetic polarization. Such bound states have been 
 observed in various systems~\cite{yazdani.jones.97,ji.zhang.08,franke.schulze.11,
ruby.pientka.15,hatter.heinrich.15,menard.guissart.15,jellinggaard.groverasmussen.16,choi.rubioverdu.17,
assouline.feuilletpalma.17}. In-gap states 
(appearing in pairs  symmetrically around the Fermi level) can be
nowadays  controlled electrostatically or magnetically 
\cite{jellinggaard.groverasmussen.16} whereas their topography, 
spatial extent and polarization can be precisely inspected by the 
state-of-art tunneling measurements~\cite{salkola.balatsky.97,flatte.byers.97}.

It has been reported that adatoms deposited on  2-dimensional  
superconducting surface develop the Yu--Shiba--Rusinov (YSR) states,
extending to a dozen of intersite distances and they reveal the 
particle-hole oscillations \cite{menard.guissart.15}. Bound states of 
these magnetic impurities in a superconducting NbSe$_{2}$ 
are characterized by the star shape~\cite{ugeda.bradley.16} typical for
the rotational symmetry of its triangular lattice. More complex objects (like dimers) reveal  
other spatial features, showing the bonding and antibonding states~\cite{kezilebieke.dvorak.17}. 
In a somewhat different context it has been pointed out \cite{Palacios-2014}
that exchange coupling between numerous quantum defects 
involving their intrinsic spins can couple them ferromagnetically, 
and this can be used (e.g. in metallic carbon nanotubes) for a robust  transmission  of magnetic information at large distances.

In all cases the bound YSR states are also sensitive to
interactions. One of them is the spin-orbit coupling 
(usually meaningful at boundaries, e.g. surfaces) 
\cite{kim.zhang.15,kaladzhyan.bena.16,Glodzik-2017}. Such interaction
in one-dimensional magnetic nanowires can induce the topologically nontrivial 
superconducting phase, in which the 
YSR states undergo mutation 
to the Majorana (zero-energy) quasiparticles. Coulomb repulsion between 
the opposite spin electrons can bring additional important effects. In the proximitized quantum dots it can lead to a parity change (quantum 
phase transition) with further influence on the subgap Kondo effect (driven by effective spin-exchange 
coupling with mobile electrons). Furthermore, such spin
exchange can be amplified by the induced electron pairing, and can have  constructive influence on the Kondo effect \cite{Zitko-2015a,Domanski-2016}.

We study here the polarized bound states, taking into account  
the spin-orbit and/or Coulomb interactions. In particular, we consider: 
(i) the single magnetic impurity in 2-dimensional square lattice of 
a superconducting host, (ii) nanoscopic chain of the magnetic impurities on 
the classical superconductor (i.e.\ proximitized Rashba nanowire) in its 
topologically trivial/nontrivial superconducting phase, and (iii) the strongly 
correlated quantum dot side-attached to the Rashba chain, where the Kondo and 
the leaking Majorana quasiparticle can be confronted with each other. 
These magnetically polarized 
YSR and 
Majorana quasiparticles as well as the subgap Kondo effect can be experimentally 
verified using the tunneling heterostructures with ferromagnetic lead (STM tip).

\section{Single magnetic impurity}
\label{sec.impurities}

Let us start by considering a single magnetic impurity on surface of the $s$-wave
superconductor in presence of the spin-orbit interactions. This situation 
can be modelled by the Anderson-type Hamiltonian
\begin{eqnarray}
\hat{\mathcal{H}} = \hat{\mathcal{H}}_{\rm sc} + \hat{\mathcal{H}}_{\rm imp} 
+  \hat{\mathcal{H}}_{\rm SOC} .
\label{eq.ham} 
\end{eqnarray}
We describe the superconducting substrate by 
\begin{eqnarray}
\hat{\mathcal{H}}_{\rm sc} = - t \sum_{ \langle i,j \rangle \sigma } \hat{c}_{i\sigma}^{\dagger} 
\hat{c}_{j\sigma} \!+\! U \sum_{i} \hat{n}_{i\uparrow} \hat{n}_{i\downarrow}
\!-\! \mu \sum_{i\sigma} \hat{n}_{i\sigma} ,
\end{eqnarray}
where $\hat{c}_{i\sigma}^{\dagger}$ ($\hat{c}_{i\sigma}$) denotes creation (annihilation) 
of electron with spin $\sigma$ at {\it i}-th site, $t$ is a hopping integral 
between the nearest-neighbors, $\mu$ is the chemical potential, and
$\hat{n}_{i\sigma}=\hat{c}_{i\sigma}^{\dagger} \hat{c}_{i\sigma}$ is the number operator.
For simplicity, we assume a weak attractive potential $U < 0$ between itinerant electrons
and treat it within the mean-field decoupling
$\hat{c}_{i\uparrow}^{\dagger} \hat{c}_{i\uparrow} \hat{c}_{i\downarrow}^{\dagger} 
\hat{c}_{i\downarrow} \approx \chi_{i} \hat{c}_{i\uparrow}^{\dagger} \hat{c}_{i\downarrow}^{\dagger} 
+ \chi_{i}^{\ast} \hat{c}_{i\downarrow} \hat{c}_{i\uparrow} - | \chi_{i} |^{2} 
+ n_{i\uparrow} \hat{c}_{i\downarrow}^{\dagger} \hat{c}_{i\downarrow} 
+ n_{i\downarrow} \hat{c}_{i\uparrow}^{\dagger} \hat{c}_{i\uparrow} 
- n_{i\uparrow} n_{i\downarrow}$,
where $\chi_{i} = \langle \hat{c}_{i\downarrow} \hat{c}_{i\uparrow} \rangle$ 
is the local superconducting order parameter and $n_{i\sigma} = \langle 
\hat{n}_{i\sigma} \rangle$. The
Hartree term can be incorporated into the local (spin-dependent) chemical 
potential $\mu \rightarrow \tilde{\mu}_{i\sigma} \equiv \mu - U n_{i\bar{\sigma}}$.
The second term in Eq.~(\ref{eq.ham}) refers to the local impurity 
\begin{equation}
\hat{\mathcal{H}}_{\rm imp} =  -J \left( \hat{c}_{0\uparrow}^{\dagger} \hat{c}_{0\uparrow}
\!-\! \hat{c}_{0\downarrow}^{\dagger} \hat{c}_{0\downarrow} \right)  
+K \left( \hat{c}_{0\uparrow}^{\dagger} \hat{c}_{0\uparrow}
\!+\! \hat{c}_{0\downarrow}^{\dagger} \hat{c}_{0\downarrow} \right)  
\label{magn_scatter}
\end{equation}
which affects the order parameter $\chi_{i}$ near the impurity  site $i\!=\!0$, inducing 
the YSR states~\cite{smith.tanaka.16,goertzen.tanaka.17}. In this work we 
focus on the magnetic term $J$~\cite{balatsky.vekhter.06,koerting.10}, 
disregarding the potential scattering $K$.

The spin-orbit coupling (SOC) can be expressed by 
\begin{eqnarray}
\hat{\mathcal{H}}_{\rm SOC} = - i \lambda \sum_{ij\sigma\sigma'} \hat{c}_{i+\bm{d}_{j}
\sigma}^{\dagger} \left( \bm{d}_{j} \times 
\hat{\bm \sigma}^{\sigma\sigma'} \right) \cdot \hat{\bm w} \; \hat{c}_{i\sigma'} ,
\end{eqnarray}
where vector ${\bm d_j} = ( d_{j}^{x} ,  d_{j}^{y} , 0 )$ refers to positions of the nearest 
neighbors of {\it i}-th site, and $\hat{\bm \sigma} = ( \sigma_{x} , \sigma_{y} , \sigma_{z} )$ 
stand for the Pauli matrices. The unit vector $\hat{\bm w}$ shows a direction of the spin orbit 
field, which  can be arbitrary. Here we restrict our considerations to the in-plane 
$\hat{\bm w} \equiv \hat{x} = ( 1 , 0 , 0 )$ polarization (that would be important for
nontrivial superconductivity in nanowires discussed in Sec.\ \ref{sec.majoranas}). The other (out-of-plane) 
component could eventually mix  $\uparrow$ and $\downarrow$ spins \cite{Glodzik-2017}.

\subsection{Bogoliubov--de Gennes technique}

Impurities break the translational invariance, therefore the pairing 
amplitude $\chi_{i}$ and occupancy $n_{i\sigma}$ have to be determined for each lattice 
site individually. We can diagonalize the Hamiltonian~(\ref{eq.ham}) by the unitary 
transformation
\begin{eqnarray}
\hat{c}_{i\sigma} = \sum_{n} \left( u_{in\sigma} \hat{\gamma}_{n} 
- \sigma v_{in\sigma}^{\ast} \hat{\gamma}_{n}^{\dagger} \right), 
\label{eq.bvtransform} 
\end{eqnarray}
where $\hat{\gamma}^{(\dagger)}_{n}$ are  quasiparticle fermionic operators, 
with eigenvectors $u_{in\sigma}$ and $v_{in\sigma}$. 
This leads to the Bogoliubov--de~Gennes (BdG) equations
\begin{eqnarray}
\label{eq.bdg} \mathcal{E}_{n} &&
\left(
\begin{array}{c}
u_{in\uparrow} \\ 
v_{in\downarrow} \\ 
u_{in\downarrow} \\ 
v_{in\uparrow}
\end{array} 
\right) \\
\nonumber = \sum_{j} && \left(
\begin{array}{cccc}
H_{ij\uparrow} & D_{ij} & S_{ij}^{\uparrow\downarrow} & 0 \\ 
D_{ij}^{\ast} & -H_{ij\downarrow}^{\ast} & 0 & S_{ij}^{\downarrow\uparrow} \\ 
S_{ij}^{\downarrow\uparrow} & 0 & H_{ij\downarrow} & D_{ij} \\ 
0 & S_{ij}^{\uparrow\downarrow} & D_{ij}^{\ast} & -H_{ij\uparrow}^{\ast}
\end{array} 
\right) 
\left(
\begin{array}{c}
u_{jn\uparrow} \\ 
v_{jn\downarrow} \\ 
u_{jn\downarrow} \\ 
v_{jn\uparrow}
\end{array} 
\right),
\end{eqnarray}
where $D_{ij} = \delta_{ij} U \chi_{i}$  and
the single-particle term is given by
$H_{ij\sigma} = - t \delta_{\langle i,j \rangle} 
- \left( \tilde{\mu}_{i\sigma}  - \sigma J  \delta_{i0} \right) \delta_{ij} 
+ S_{ij}^{\sigma\sigma}$ 
with the spin-orbit coupling term
$S_{ij}^{\sigma\sigma'} = - i \lambda  \sum_{l} \left( {\bm d}_{l} \times 
\hat{\bm \sigma}^{\sigma\sigma'} \right) \cdot \hat{\bm w} \; \delta_{j,i+{\bm d}_{l}}$.
Here, $S_{ij}^{\sigma\sigma}$ and $S_{ij}^{\sigma\bar{\sigma}}$ (where $\bar{\sigma}$ is opposite to $\sigma$) correspond to in--plane and out--of--plane spin orbit field, respectively, which satisfy $S_{ij}^{\sigma\sigma'} = ( S_{ji}^{\sigma'\sigma} )^{\ast}$. 

Solving numerically the BdG equations~(\ref{eq.bdg}) we can determine 
the local order parameter $\chi_{i}$ and occupancy $n_{i\sigma}$ 
\begin{eqnarray}
\chi_{i} &=&  
\sum_{n} \left[ u_{in\downarrow} v_{in\uparrow}^{\ast} f( \mathcal{E}_{n} ) 
- u_{in\uparrow} v_{in\downarrow}^{\ast} f ( - \mathcal{E}_{n} ) \right] , \\
n_{i\sigma} &=&  
\sum_{n} \left[ | u_{in\sigma} |^{2} f( \mathcal{E}_{n} ) + 
| v_{in\bar{\sigma}} |^{2} f ( - \mathcal{E}_{n} ) \right] ,
\end{eqnarray}
where $ f ( \omega ) = \left[ 1 + \exp ( \omega / k_{B} T ) \right]^{-1}$. 
In what follows, we shall inspect the spin-resolved local density of states 
\begin{eqnarray}
\rho_{i\sigma} ( \omega ) = \sum_{n} \left[ | u_{in\sigma} |^{2} \delta 
( \omega - \mathcal{E}_{n} ) + | v_{in\sigma} |^{2} \delta ( \omega + \mathcal{E}_{n} ) 
\right] . 
\nonumber
\end{eqnarray}
For its numerical computation we replace the Dirac 
delta function by Lorentzian $\delta (\omega) = \zeta / [ \pi ( \omega^{2} 
+ \zeta^{2})]$ with a small broadening $\zeta =0.025 t$.
We have solved the BdG equations, considering the single magnetic impurity in 
a square lattice, comprising $N_{a} \times N_{b} = 41 \times 41$ sites.
We assumed $U/t = - 3$, $\mu/t = 0$, and determined the bound states 
for two representative values of the spin-orbit coupling $\lambda$ upon 
varying $J$.

\subsection{Topography of the bound states}
\label{sec.topography}

\begin{figure}[!t]
\centering
\includegraphics[width=0.9\linewidth]{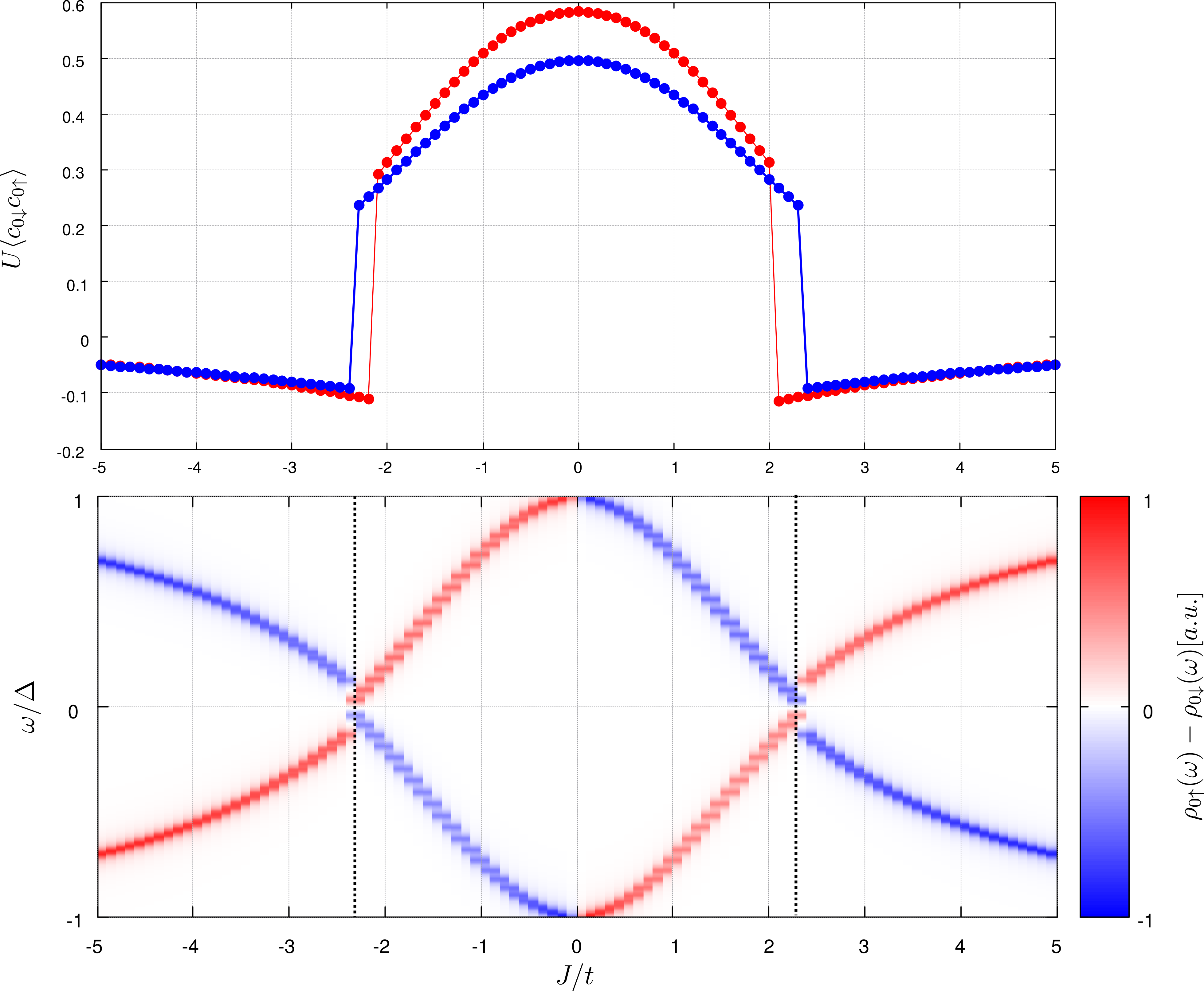}
\caption{
The local order parameter 
obtained at zero temperature for the weak $\lambda/t=0.1$ (red line) and strong 
spin orbit coupling $\lambda/t=1$ (blue line). The bottom panel shows 
the energies and magnetic polarization $\rho_{0\uparrow}(\omega) - \rho_{0\downarrow}
(\omega)$ of YSR states obtained in the weak coupling limit $\lambda/t = 0.1$.
\label{fig.qcp}
}
\end{figure}

The magnetic potential has substantial influence on the local order parameter 
$\chi_0$. In particular, at some critical value $J_{c}$ this quantity discontinuously 
changes its magnitude and sign (see the upper panel in Fig.~\ref{fig.qcp}), signalling 
the first-order phase transition~\cite{pershoguba.bjornson.15,glodzik.ptok.17,
mashkoori.bjornson.17}. This quantum phase transition at $J_{c}$ is an artifact 
of the classical spin approximation. When spin fluctuations are allowed, 
a Kondo-like crossover is obtained instead of a first-order phase 
transition~\cite{Satori.92,Sakai.93}.
%
%
In general, the quasiparticle spectrum at the impurity site is characterized 
by two bound states $\pm E_{\rm YSR}$ inside the gap $\Delta$ of 
superconducting host (displayed in bottom panel of Fig.\ \ref{fig.qcp}). 
These energies $\pm E_{\rm YSR}$ and the related spectral weights depend on $J$. 
At $J=J_{c}$ the YSR bound states cross each other $E_{\rm YSR}(J_{c})\!=\!0$ 
and their crossing signifies the ground state parity change~\cite{sakurai.70} 
from the BCS-type (spinless) to the singly occupied (spinful) configurations 
\cite{salkola.balatsky.97,franke.schulze.11,kaladzhyan.bena.16,vangervenoei.tanaskovic.17}.
Let us remark that this quantum phase transition is also accompanied with 
reversal of the YSR polarization (see bottom panel in Fig.~\ref{fig.qcp}). Similar
behavior can be observed also for the multiple impurities, at several critical
values of $J$~\cite{morr.yoon.06}.

\begin{figure}[!t]
\centering
\includegraphics[width=0.8\linewidth]{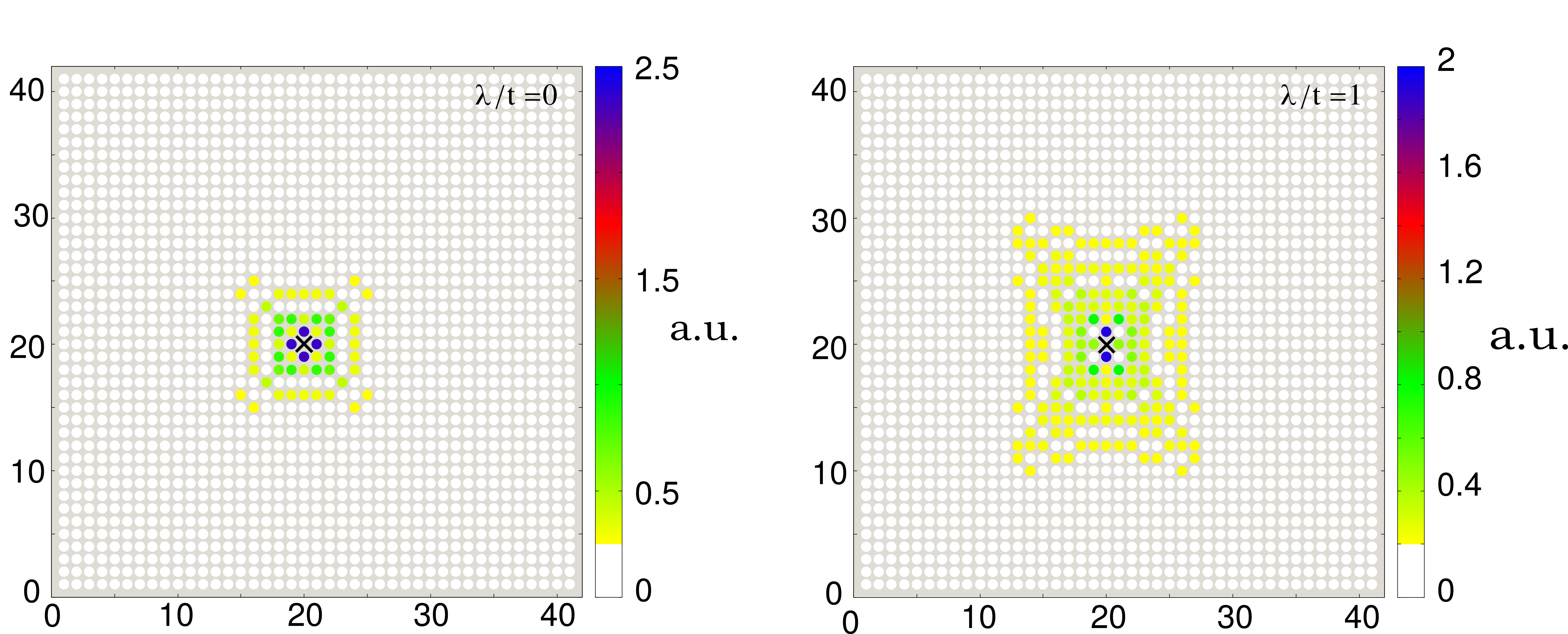}
\caption{Spatial profiles of the YSR states $\sum_{\sigma}\rho^{+}_{i\sigma}$
obtained for $|J|<J_{c}$ in absence of the spin-orbit coupling (left panel) 
and for the strong  in-plane coupling $\lambda=t$ (right panel). The spin-orbit 
field is choosen along $x$ axis and leads to additional imaginary hopping  
term along $y$ axis, which elongates of the YSR states in $y$ direction. 
The impurity spin is oriented along $(0,0,1)$ direction.
\label{fig.one_dos}
}
\end{figure}
 
Within the BdG approach we can inspect spatial profiles of the YSR 
states by integrating the spectral weights $\rho_{i\sigma}^{\pm} = 
\int_{\omega_{1}}^{\omega_{2}} \rho_{i\sigma}(\omega) \; d\omega$
in the interval $\omega \in (\omega_{1}, \omega_{2})$ capturing 
the quasiparticles at negative/positive energies $\pm E_{\rm YSR}$~\cite{rontynen.ojanen.15}.
Fig.\ \ref{fig.one_dos} illustrates the results obtained for $\lambda=0$ 
(left panel) and $\lambda=t$ (right panel). 
We clearly notice a fourfold  rotational symmetry (typical for
the square lattice) and the spatial extent of YSR states reaching several sites 
away from the magnetic impurity.
Nonvanishing difference of the spectral weights $|u_{in\uparrow}|^{2}-|u_{in\downarrow}|^{2}$
at the positive energy $\omega=+E_{YSR}$ and $|v_{in\uparrow}|^{2}-|v_{in\downarrow}|^{2}$ 
at the negative energy $\omega=-E_{YSR}$ implies effective 
spin-polarization of the bound states
(their polarization is illustrated in the bottom panel of Fig.\ 1).

\begin{figure}[!t]
\centering
\includegraphics[width=0.7\linewidth]{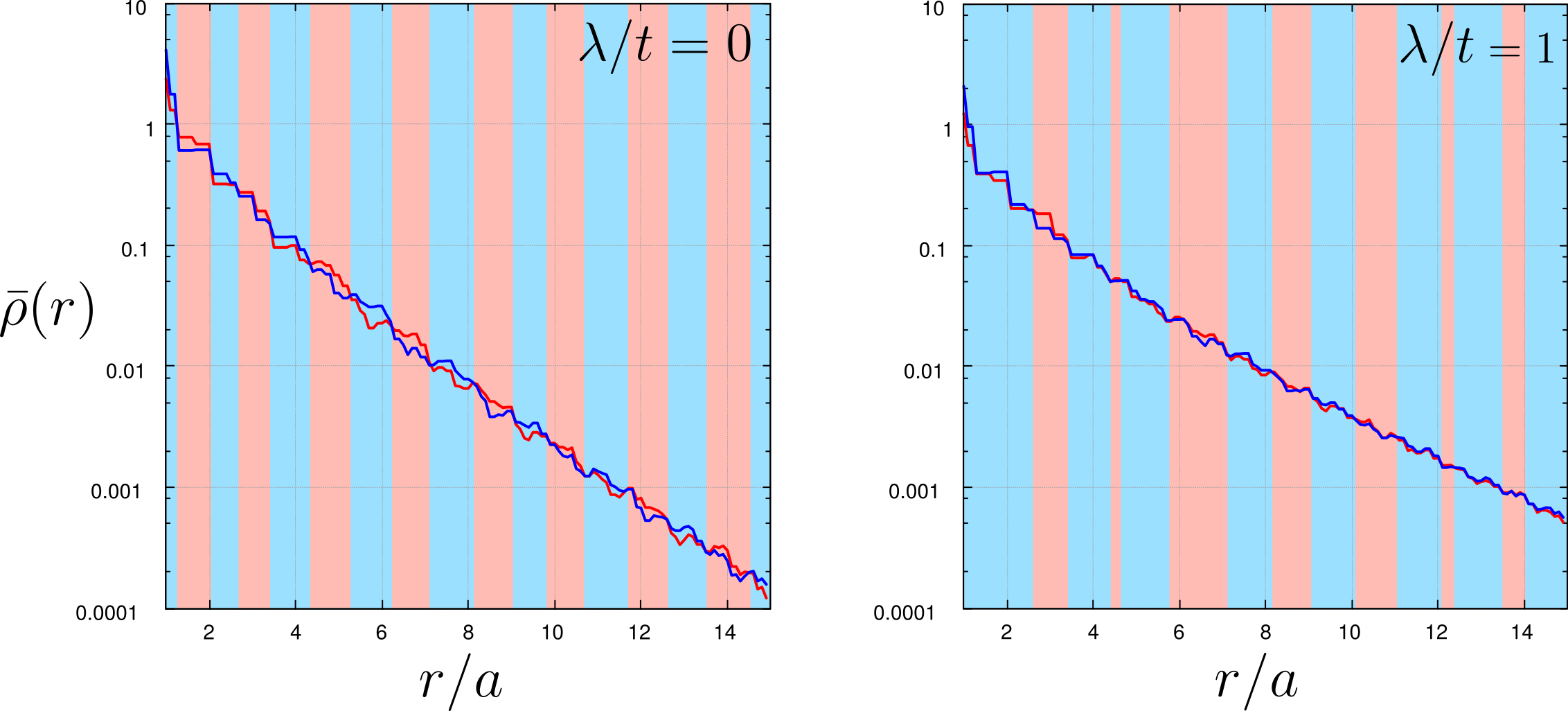}
\caption{
Hole- (blue line) and electron-like (red line) displaced 
moving average $\bar{\rho}^{\pm} ( r )$ as a function of radial distance $r$ 
from the impurity site obtained for $| J | < J_{c}$ using $\delta r = 0.5a$. 
The (blue/red) background color indicates the dominant hole/particle 
type of YSR state at a given distance $r$. The upper and lower panels 
correspond to $\lambda=0$ and  $\lambda=t$, respectively.
\label{fig.ph_oscillation}
}
\end{figure}

For some quantitative estimation of the spatially varying magnetization 
(driven by the particle-hole asymmetry) we have computed the {\em displaced 
moving average} $\bar{\rho}^{\pm}(r)$, corresponding to an averaged spectral 
weight contained in a ring of the radius $r$ and a small half-width $\delta r$. 
This quantity is sensitive only to radial distance $r$ from the magnetic 
impurity, averaging the angular anisotropy. Our results, presented in 
Fig.~\ref{fig.ph_oscillation}, clearly indicate the spatial particle-hole oscillations
$\bar{\rho}^{\pm}(r)$ of the YSR states (compare the blue and red lines). 
Such particle-hole oscillations decay exponentially with $r$ in agreement 
with previous studies~\cite{morr.stavropoulos.03,kawakami.hu.15,menard.guissart.15}. 
The dominant (particle or hole) contributions to the YSR bound states are displayed
by an alternating color of the background in Fig.~\ref{fig.ph_oscillation}.
We notice that the  spin-orbit coupling seems to suppress these
particle-hole oscillations.

Summarizing this section, we point out that the quantum phase transition 
(at $J_{c}$) depends on the spin-orbit coupling $\lambda$ and it has 
experimentally observable consequences in the magnetization induced near 
the impurity site. For weak magnetic scattering $| J | < J_{c}$ the impurity 
is partly screened, whereas for the stronger couplings $| J | > J_{c}$ 
the impurity polarizes its neighborhood in a direction of its own magnetic 
moment. Similar effects have been previously discussed by V.~Kaladzhyan 
{\em et al.}~\cite{kaladzhyan.bena.16} but here we additionally
present the role of spin orbit coupling. First of all, such interaction
shifts the quantum phase transition (to larger values of $J$) and secondly
it enhances the spatial extent of YSR states and gradually smoothens
their particle-hole oscillations.


\section{Magnetically polarized Majorana quasiparticles}
\label{sec.majoranas}

In this section we increase the number of impurities.
Let us now imagine the nanoscopic chain of magnetic impurities (for instance
Fe atoms) deposited on a surface of the conventional $s$-wave superconductor.
We shall study the magnetically polarized bound states, focusing on the
proximity induced nontrivial superconducting phase. In practice, the 
quasiparticle spectrum can be probed within STM-type setup, by attaching
the conducting  \cite{Yazdani-14,Kisiel-15}, superconducting \cite{Franke-15},
or the magnetically polarized tip \cite{Yazdani-2017}. We shall assume the 
spin-orbit interaction perpendicularly aligned to the wire and magnetic
field parallel to it, leading to the effective intersite pairing 
of identical spins and (under specific condition) inducing the zero-energy
end modes resembling the Majorana quasiparticles. This issue has been
recently studied very intensively but here we simply focus on the spin-polarized
aspects of this problem.

Due to the spin-orbit interaction, momentum and spin are no longer ``good''
quantum numbers. By solving the problem numerically, however, we can estimate 
percentage with which the true quasiparticles are represented by the initial
spin. We have recently emphasized \cite{Maska-2017}, that 
amplitude of the intersite pairing (between identical spin electrons) differs 
several times for $\uparrow$ and $\downarrow$ sectors. This 
leads to an obvious polarization of the 
YSR and Majorana 
quasiparticles (the latter appearing near the nanochain edges).

\subsection{Proximitized Rashba chain}

Let us consider the STM-type geometry, relevant to the recent experimental 
situation addressed by A. Yazdani and coworkers \cite{Yazdani-2017}, which 
can be described by the following Hamiltonian
\begin{eqnarray} 
\hat{\mathcal{H}} &=& \hat{\mathcal{H}}_{\rm tip} + \hat{\mathcal{H}}^{\rm prox}_{\rm chain}  
+  \hat{\mathcal{H}}_{\rm tip-chain} .
\label{model}
\end{eqnarray} 
We assume here, that STM tip describes the polarized fermion gas 
$\hat{\mathcal{H}}_{\rm N} \!=\! \sum_{{\bf k},\sigma} \xi^{\sigma}_{{\bf k}{\rm N}}  
\hat{c}_{{\bf k} \sigma {\rm N}}^{\dagger} \hat{c}_{{\bf k} \sigma {\rm N}}$,   
where energy $\xi^{\sigma}_{{\bf k}{\rm N}}=\varepsilon_{\bf k}
-\mu_{{\rm N}\sigma}$ can be controlled by some finite detuning of the chemical 
potentials $\mu_{{\rm N}\uparrow}-\mu_{{\rm N}\downarrow}$. 
Individual atoms of the nanochain are coupled with such
STM tip via $\hat{\mathcal{H}}_{\rm tip-chain} 
= \sum_{{\bf k},\sigma}  \left( V_{i,{\bf k} {\rm N}} \; 
\hat{d}_{i,\sigma}^{\dagger}  \hat{c}_{{\bf k} \sigma {\rm N}} 
+  V_{i,{\bf k} \beta}^{*}  \hat{c}_{{\bf k} \sigma {\rm N}}^{\dagger} 
\hat{d}_{i,\sigma} \right)$. For simplicity, we assume constant
couplings $\Gamma_{\beta}=2\pi \sum_{\bf k} 
|V_{i,{\bf k}\beta}|^2 \; \delta(\omega \!-\! \xi_{{\bf k}\beta})$.

\begin{figure} 
\includegraphics[width=0.95\linewidth]{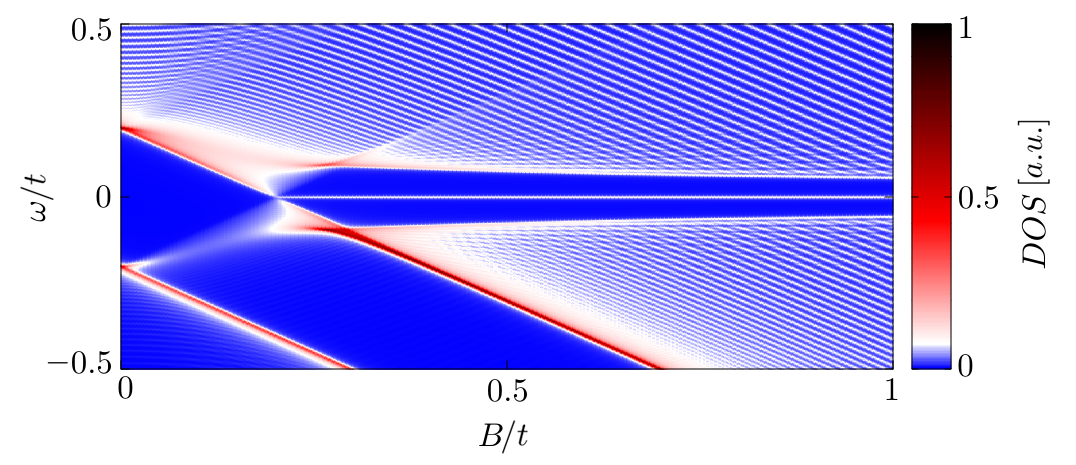}
\includegraphics[width=0.95\linewidth]{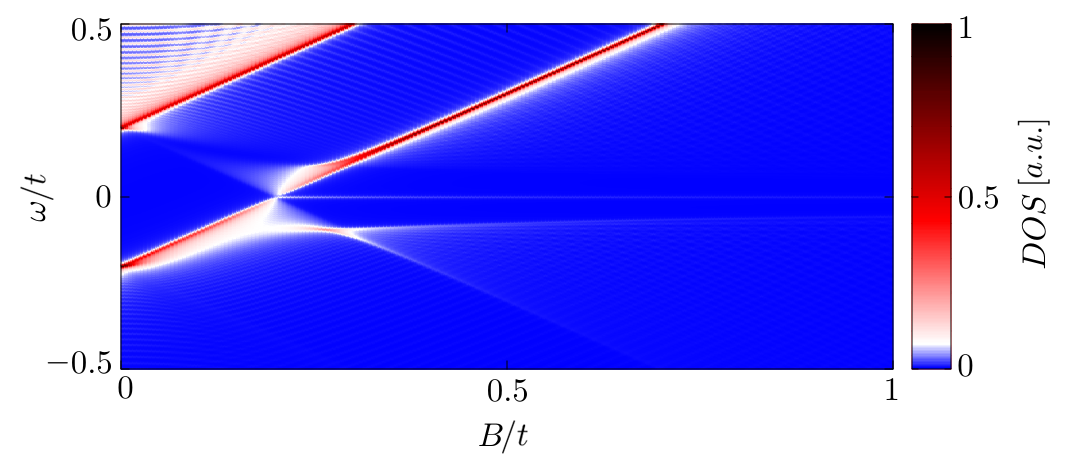}
\caption{The effective quasiparticle spectrum $\rho_{i,\sigma}(\omega)$
with respect to the magnetic field $B$ aligned along the nanochain
obtained for $\sigma\!=\:\uparrow$ (upper panel)
and $\sigma\!=\:\downarrow$ (bottom panel).
Magnetic field $B$ is expressed in units of $t/(g\mu_{B}/2)$.}
\label{Rashba}
\end{figure}

Low-energy physics of such proximitized Rashba 
nanowire can be described by \cite{DasSarma-2016}
\begin{eqnarray}
\hat{\mathcal{H}}^{\rm prox}_{\rm chain} &=& 
\sum_{i,j,\sigma} \left( t_{ij} -\delta_{ij} \mu\right) 
\hat{d}^{\dagger}_{i,\sigma} \hat{d}_{j,\sigma} 
+ \hat{\mathcal{H}}_{\rm Rashba} 
\nonumber \\ &+& 
 \hat{\mathcal{H}}_{\rm Zeeman} + \hat{\mathcal{H}}_{\rm prox},
\label{chain_model}
\end{eqnarray}
where $\hat{d}_{i,\sigma}^{(\dag)}$ annihilates (creates) electron 
of spin $\sigma$ at site $i$ with energy $\varepsilon_{i}$ and 
$t_{ij}$ is the hopping integral. The effective intersite
($p$-wave) pairing is induced via combined effect of the Rashba 
and the Zeeman terms
\begin{eqnarray}
    \hat{\mathcal{H}}_\mathrm{Rashba}&=&-\alpha\sum_{i,\sigma,\sigma'}\left[
    \hat{d}^{\dagger}_{i+1,\sigma}
    \left(i\sigma^y\right)_{\sigma\sigma'}\hat{d}_{i,\sigma'}+\mathrm{H.c.}\right],\\
    \hat{H}_\mathrm{Zeeman}&=&\frac{g\mu_\mathrm{B}B}{2}\sum_{i,\sigma,\sigma'}
    \hat{d}^{\dagger}_{i,\sigma}\left(\sigma^z\right)_{\sigma\sigma'}
    \hat{d}_{i,\sigma'} .
\end{eqnarray}
The  proximity effect, which induces the on-site (trivial) pairing, 
can be modelled as \cite{Stanescu-13}
\begin{eqnarray} 
\hat{\mathcal{H}}_{\rm prox}= \Delta_{i} \left( \hat{d}_{i,\uparrow}^{\dagger} 
\hat{d}_{i,\downarrow}^{\dagger} + \hat{d}_{i,\downarrow}
\hat{d}_{i,\uparrow} \right) 
\label{large_Delta} 
\end{eqnarray} 
with the local pairing potential $\Delta_{i}=\Gamma_{\rm S}/2$.

Fig.\ \ref{Rashba} shows evolution of the spin-dependent spectrum
$\rho_{i,\sigma}(\omega)$ with respect to varying magnetic
field. At critical value ($B \simeq 0.2$) we observe emergence of
the zero-energy quasiparticles, whose spectral weights strongly
depend on the spin $\sigma$. 

\subsection{Spin-polarized Majorana quasiparticles}

%
\begin{figure} 
\includegraphics[width=\linewidth]{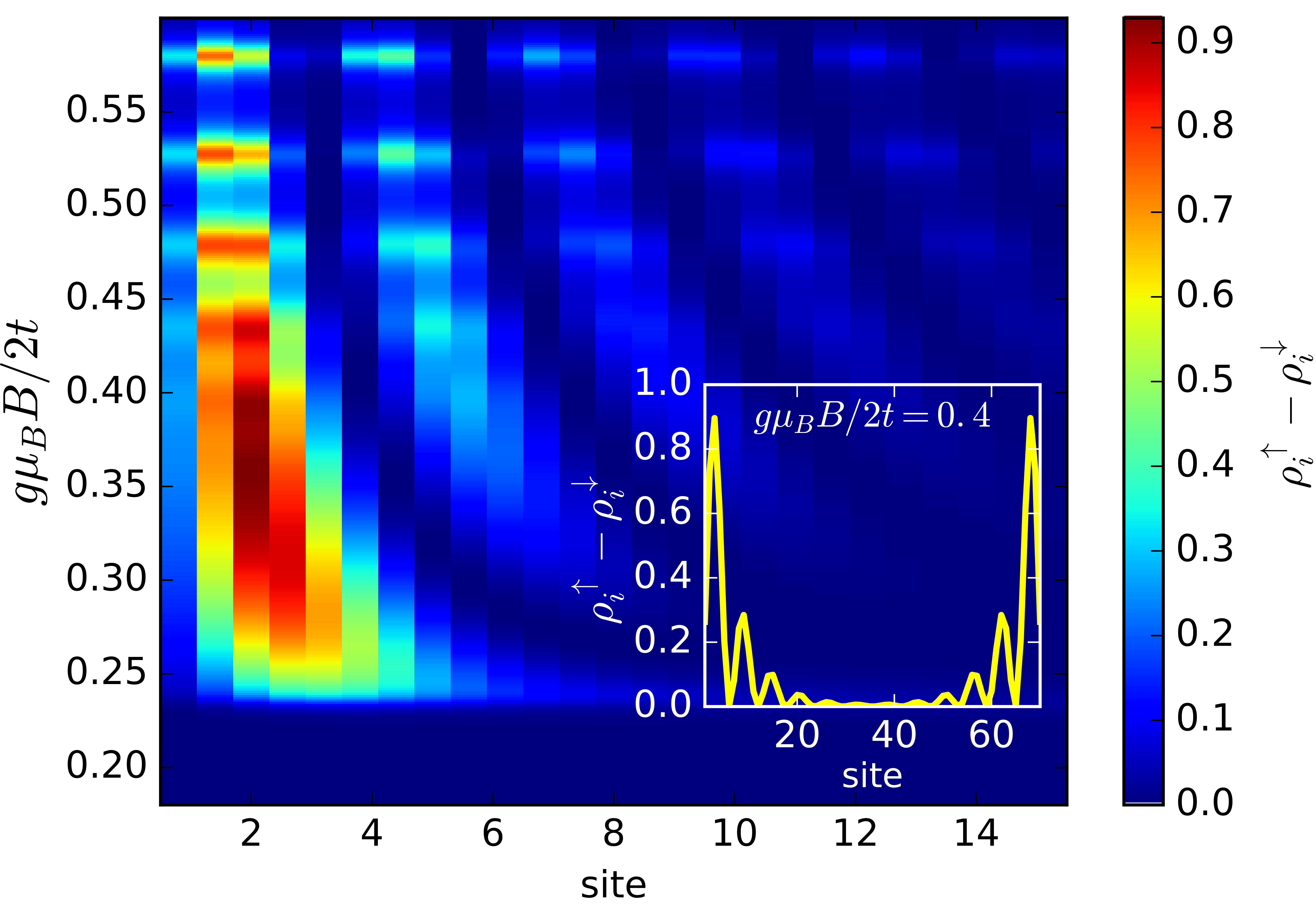}
\caption{Magnetically polarized spectrum $\rho_{i,\uparrow}(\omega)
-\rho_{i,\downarrow}(\omega)$  at $\omega=0$ obtained at peripherial 
sites of the  Rashba chain.}
\label{oscillations}
\end{figure}

For  better understanding of the polarized zero-energy quasiparticles,
we present in  Fig.\ \ref{oscillations} the spatial profiles of the zero-energy 
(Majorana) quasiparticles. As usually such quasiparticles emerge near
the edges on a nanoscopic chain, practically over 10 to 15 sites (see inset). 
Let us notice the substantial quantitative difference between these 
zero-energy quasiparticles appearing in $\uparrow$ and $\downarrow$ spin 
sectors. Such `intrinsic polarization' of the Majorana modes has been 
previously suggested in Ref.\ \cite{Striclet-2012}, and recently we have
proposed \cite{MaskaDomanski-2017} their empirical detection by means 
of the  {Selective Equal Spin Andreev Reflection} (SESAR) spectroscopy.

The main idea is to apply  bias voltage $V$ between the STM tip 
and the superconducting substrate, inducing the charge transport which 
in a subgap regime ($|V| \ll \Delta/|e|$) originates from the Andreev 
(particle to hole) scattering mechanism. The polarized Andreev current 
can be expressed  by Landauer-B\"uttiker formula
\begin{eqnarray} 
I_{i}^{\sigma}(V) = \frac{e}{h} \int \!\!  d\omega \; T_{i}^{\sigma}(\omega)
\left[ f(\omega\!-\!eV)\!-\!f(\omega\!+\!eV)\right] ,
\label{I_A}
\end{eqnarray} 
where the transmittance 
$T_{i}^{\sigma}(\omega) =  \Gamma_{\rm N}^{2}   \left| \langle\langle 
\hat{d}_{i\sigma} \hat{d}_{{i+1}\sigma}\rangle\rangle \right|^{2}
+ \Gamma_{\rm N}^{2} \left| \langle\langle \hat{d}_{i\sigma} \hat{d}_{{i-1}\sigma}
\rangle\rangle \right|^{2} $ 
and 
$T_{1}^{\sigma}(\omega) =  \Gamma_{\rm N}^{2} \; \left| \langle\langle 
\hat{d}_{1\sigma} \hat{d}_{{2}\sigma}\rangle\rangle \right|^{2}$,
$T_{\rm N}^{\sigma}(\omega) =  \Gamma_{\rm N}^{2} \; \left| \langle\langle 
\hat{d}_{N\sigma} \hat{d}_{{N-1}\sigma}\rangle\rangle \right|^{2}$. 
The anomalous Green's functions can be computed numerically from the solution 
of the Bogoliubov--de Gennes equations of this model (\ref{chain_model}).
The net spin current $I_{i}^{\rm spin}(V) = I_{i}^{\uparrow}(V)-I_{i}^{\downarrow}(V)$
turns out to be predominantly sensitive to the Majorana end-modes. Its 
differential conductance $G_{i}^{\rm spin}(V) = \frac{ \partial }{\partial V}
I^{\rm spin}_{i} (V)$ can thus distinguish the polarized Majorana quasiparticle
(near $V=0$) from the 
YSR states (appearing at finite voltage).

\subsection{Leakage of polarized Majorana quasiparticles}

\begin{figure} 
\includegraphics[width=0.9\linewidth]{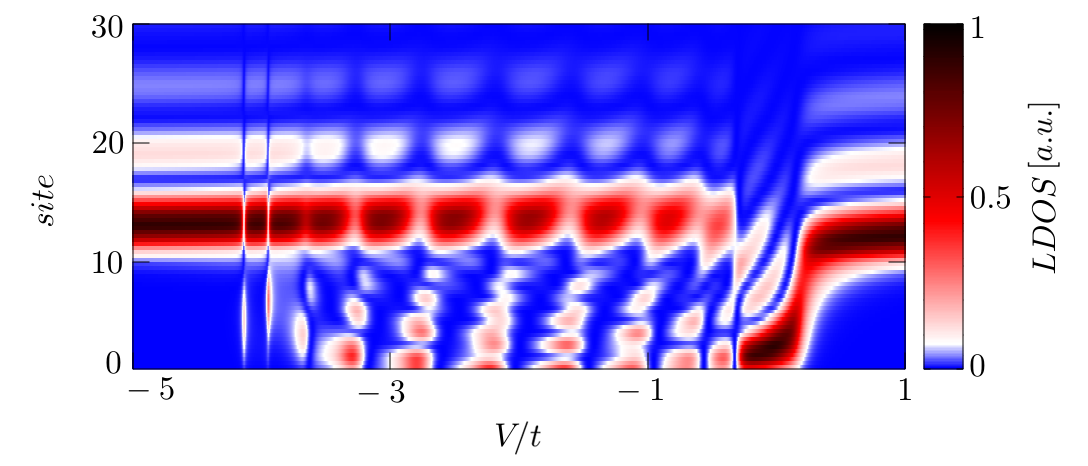}
\includegraphics[width=0.9\linewidth]{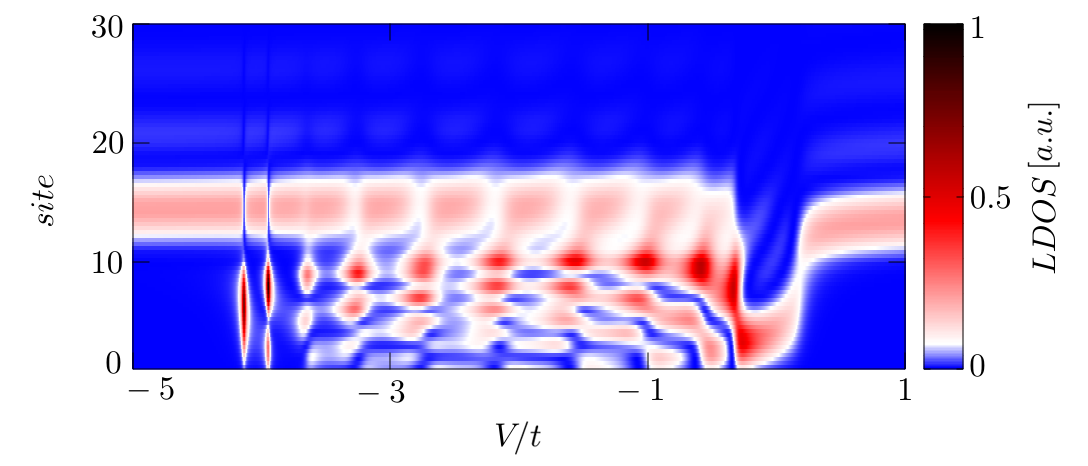}
\caption{Leakage of the spin-polarized  Majorana quasiparticles 
from the topological superconducting phase  of the Rashba chain 
($i\geq 10$) onto the side-attached multi-site ($i \in \left< 1 ;  10 \right>$) quantum
dot. The upper and bottom panel shows $\rho_{i,\sigma}(\omega)$ for $\omega\!=\!0$ 
of $\uparrow$ and $\downarrow$ spin, respectively.}
\label{proxy}
\end{figure}

Bound states can leak to other side-attached nanoscopic objects. Such
proximity effect has been also predicted for the Majorana quasiparticles
by E. Vernek {\em et al.} \cite{Vernek-2014} and it has been indeed observed 
experimentally by M. T. Deng {\it et al.}~\cite{Deng-2017}. Inspired by this
achievement, there has been intensive study of the 
YSR states  coalescencing into the zero-energy Majorana state in the side-coupled
quantum dots driven by electrostatic or magnetic fields 
\cite{DasSarma-2017,Klinovaja-2017,Kobialka-2017}. 
This issue would be particularly important when attempting to braid
the Majorana end modes, e.g., in T-shape nanowires upon turning on/off 
the topological superconducting phase in its segments. We briefly 
analyse here the polarized zero-energy Majorana modes leaking 
on the multi-site quantum dot (comprised of 10 lattice sites)
side-attached to the proximitized Rashba chain (discussed above).

Fig.~\ref{proxy} displays spatial profile of the polarized spectrum 
obtained at $\omega=0$ as function of the gate voltage $V_{g}$,
which detunes the energies $V_{g}=\epsilon_{i}-\mu$ of the multi-site
($1\leq i \leq 10$) quantum dot. For numerical calculations we used 
the model parameters $\lambda=0.15 t$, $\mu=-2t$, $\Delta_{i}=0.2t$ 
and $B > B_{c}$, which guarantee the Rashba chain to be in its topologically 
nontrivial superconducting phase, hosting the zero-energy Majorana 
quasiparticles (intensive black or red regions). We clearly observe 
that for some values of $V_{g}$ these Majorana modes spread over
the entire quantum dot region. By inspecting Fig.~\ref{proxy} we furthermore
notice the pronounced spatial oscillations of these zero-energy modes.
In our opinion, this is a signature of a partial  delocalization
of the polarized Majorana quasiparticles. Surprisingly, this process
seems to be less efficient in the minor spin ($\sigma=\:\downarrow$) section.
Such effect has to be taken into account, when designing  nanostructures 
for a controllable spatial displacement of the Majorana modes (criticial for 
realization of quantum computations with use of the Majorana-based qubits)
either by electrostatic or magnetic means. Some proposals for such 
nanodevices have been recently discussed by several authors 
\cite{Kobialka-2017,chevallier.szumniak.17}. 

\vspace{0.2cm}
In summary of this section, we emphasize that the Majorana modes
coalescing from the 
YSR states in the proximitized
Rashba nanowire are characterized by their magnetic polarization. 
Indeed, such feature has been recently observed by STM spectroscopy 
with use of the polarized tip \cite{Yazdani-2017}. We have studied
here evolution of the polarized quasiparticle states with respect to
the magnetic field (Fig. \ref{Rashba}) and investigated the spatial 
oscillations of the Majorana zero-energy modes near the chain edges 
(Fig. \ref{oscillations}). Finally, we analyzed leakage of
the polarized Majorana modes on the multi-site quantum dots, 
revealing their partial delocalization (Fig. \ref{proxy}).

\section{Majorana vs Kondo effect}
\label{sec.SOPT}

\begin{figure}[t]
\centering
\includegraphics[width=0.6\linewidth]{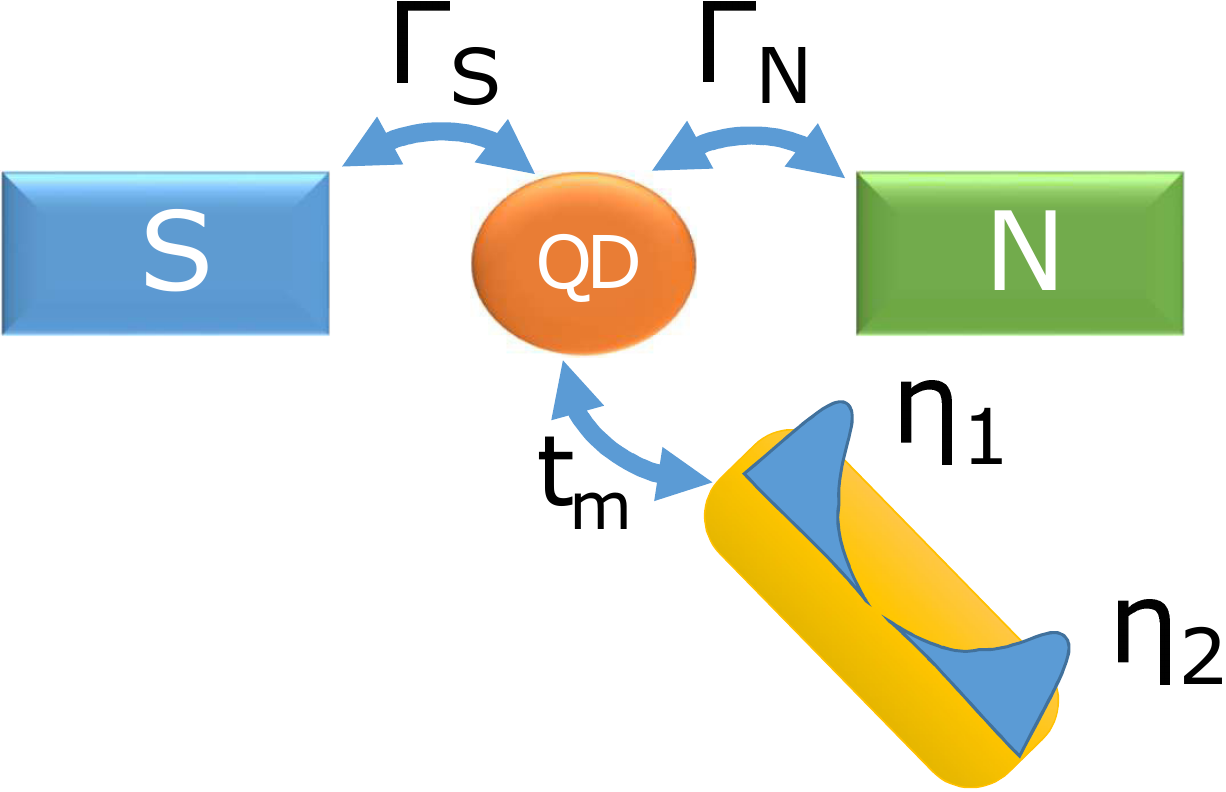}
\caption{Schematic illustration of the quantum dot (QD) coupled between
the metallic (N) and superconducting (S) leads and hybridized with 
the Rashba nanowire, hosting the Majorana quasiparticles 
$\eta_{1}$ and $\eta_{2}$ at its edges. }
\label{schematics}
\end{figure}

In Sec.\ \ref{sec.majoranas} we have discussed the polarized Majorana 
modes leaking on side-attached objects, like single impurities or  
segments of the normal nanowires. In this section we shall focus
on the correlation effects  \cite{Chirla-2016,Prada-2017,Baranski-2017},  
confronting the Majorana quasiparticle with the Kondo effect (both 
manifested at zero energy). This can be practically achieved using 
STM-type configurations sketched in Fig.\ \ref{schematics}. In 
particular we consider the subgap Kondo effect, effectively driven 
by the Coulomb repulsion $U$ and coupling of the quantum dot (QD)
with the normal lead $\Gamma_{\rm N}$ in presence of the electron 
pairing (induced via $\Gamma_{\rm S}$), which plays a prominent influence
on the spin-polarized bound states of QD. Basic mechanism of
this subgap Kondo effect showing up near the quantum phase transition 
has been earlier considered by us in absence of the Rashba nanowire 
\cite{Domanski-2016,Domanski-2017}. Our considerations can be
practically verified within STM geometry \cite{Yazdani-14,Kisiel-15} 
using the magnetic atoms (e.g. Fe) and the side-coupled nonmagnetic 
atoms (for instance Ag or Au) deposited on the superconducting 
substrate (like Pb or Al) and probed the conducting STM tip 
\cite{Yazdani-2017}. 

\subsection{Low energy model}
\label{sec.Kondomodel}

Topological superconducting phase, hosting the  Majorana modes, 
can be driven in semiconducting wires \cite{Mourik-12,Kouwenhoven-2016} 
or in nanochains of magnetic atoms 
\cite{Yazdani-14,Kisiel-15,Franke-15,Yazdani-2017} due to the nearest 
neighbor equal spin pairing. Efficiency of such $p$-wave pairing 
differs for each spin \cite{MaskaDomanski-2017}, giving rise to
polarization of the Majorana quasiparticles, with noticeable preference 
for $\uparrow$ sector (see Fig. \ref{Rashba}). In order to study 
the correlation effects we shall assume here a complete polarization 
of such Majorana quasiparticles. We thus focus, for simplicity, on 
the topological state originating from intersite pairing of only $\uparrow$ 
electrons and consider its interplay with the correlations. Let us 
remark, however, that superconducting lead  mixes both the QD spins 
with the side-attached Majorana quasiparticle \cite{Golub-2015}. 
In consequence we shall observe an interesting and spin-dependent 
relationship between the Majorana and Kondo states which could be 
probed by the polarized Andreev (particle to hole conversion) mechanism.

Our setup (Fig.\ \ref{schematics}) can be  
described by the following Anderson-type Hamiltonian
\begin{eqnarray}
\hat{\mathcal{H}} = \sum_{\beta={\rm S,N}} \left( \hat{\mathcal{H}}_{\beta} 
+ \hat{\mathcal{H}}_{\beta - {\rm QD}} \right) 
+\hat{\mathcal{H}}_{\rm QD} +\hat{\mathcal{H}}_{\rm MQD} ,
\label{HAnd}
\end{eqnarray}
where $\hat{\mathcal{H}}_{\rm N}$ corresponds to the metallic electrode,  
$\hat{\mathcal{H}}_{\rm S}$ refers to the $s$-wave superconducting substrate
and the correlated QD is modeled  by $\hat{\mathcal{H}}_{\rm QD}=\sum_{\sigma} 
\epsilon \hat{d}^{\dagger}_{\sigma} \hat{d}_{\sigma}+U\hat{n}_{\downarrow} 
\hat{n}_{\uparrow}$, where $\epsilon$ denotes the energy level and $U$ stands 
for the repulsive interaction between opposite spin electrons.  Such 
QD is coupled to both $\beta={\rm N,S}$ reservoirs via $\hat{
\mathcal{H}}_{\beta - {\rm QD}}=\sum_{{\bf k},\sigma}(V_{{\bf k}\beta}
\hat{d}^{\dagger}_{\sigma} \hat{c}_{{\bf k}\sigma \beta}+{\rm H.c.})$ and we 
assume a wide bandwidth limit, using the constant couplings $\Gamma_{\beta}$. 
It can be shown \cite{Bauer-2007,Yamada-2011,Rodero-2011,Baranski-2013} 
 that for energies $|\omega| \ll \Delta$ the super\-conducting electrode  
induces the static on-dot pairing $\hat{\mathcal{H}}_{\rm S}+\hat{\mathcal{H}}_{\rm S-QD} 
\approx H_{\rm prox}=\sum_{\sigma} \epsilon \hat{d}^{\dagger}_{\sigma} \hat{d}_{\sigma}
+U \hat{n}_{\downarrow}\hat{n}_{\uparrow} - \frac{\Gamma_{\rm S}}{2}(\hat{d}_{\uparrow} 
\hat{d}_{\downarrow}+\hat{d}^{\dagger}_{\downarrow} \hat{d}^{\dagger}_{\uparrow})$. 
One can  take into account the finite 
magnitude of superconducting gap  \cite{DasSarma-2017} but this does not
affect the main conclusions of our study.

The effective Majorana modes of the nanowire can be modeled 
by  \cite{Lutchyn-2015} $\hat{\mathcal{H}}_{\rm MQD} = i\epsilon_m 
\hat{\eta}_{1} \hat{\eta}_{2} + \lambda ( \hat{d}_{\uparrow} \hat{\eta}_{1} 
+  \hat{\eta}_{1} \hat{d}^{\dagger}_{\uparrow})$,  where
$\hat{\eta}_{i}=\hat{\eta}_{i}^{\dagger}$ are hermitian operators 
and $\epsilon_{m}$ corresponds to an overlap between Majoranas. 
We recast these operators by the standard fermionic ones \cite{Franz-15}
$\hat{\eta}_{1}=\frac{1}{\sqrt{2}}(\hat{f}+\hat{f}^{\dagger})$ and
$\hat{\eta}_{2}=\frac{-i}{\sqrt{2}}(\hat{f}-\hat{f}^{\dagger})$.
Finally, the Hamiltonian (\ref{HAnd}) simplifies to
\begin{eqnarray}
\hat{\mathcal{H}} &=& 
\hat{\mathcal{H}}_{\rm N} + \hat{\mathcal{H}}_{\rm N - QD} +
\sum_{\sigma} \epsilon \hat{d}^{\dagger}_{\sigma} \hat{d}_{\sigma}
+U \hat{n}_{\downarrow}\hat{n}_{\uparrow} 
- \frac{\Gamma_S}{2}(\hat{d}_{\uparrow} \hat{d}_{\downarrow}
 \nonumber \\ 
&+& \hat{d}^{\dagger}_{\downarrow} \hat{d}^{\dagger}_{\uparrow})
+ \epsilon_m \hat{f}^{\dagger} \hat{f} 
+ t_m (\hat{d}^{\dagger}_{\uparrow} - \hat{d}_{\uparrow}) ( \hat{f} + 
\hat{f}^{\dagger} ) - \frac{\epsilon_m}{2}  
\end{eqnarray}
with the auxiliary coupling $t_{m}=\lambda/\sqrt{2}$. The subgap 
Kondo physics originates in this model from the Coulomb term 
$U \hat{n}_{\downarrow}\hat{n}_{\uparrow}$ and the effective
spin exchange interactions due to $\hat{\mathcal{H}}_{\rm N - QD}$.
It has been shown \cite{Zitko-2015a,Domanski-2016} that under
specific conditions the on-dot pairing can cooperate with 
the subgap Kondo effect. This particular situation
occurs only near the quantum phase transition.

\subsection{Spin-dependent spectrum}
\label{sec.Kondo}

Let us examine how the subgap Kondo effect gets along with 
the Majorana mode. Earlier studies of the correlated quantum dot 
coupled to both normal (conducting) electrodes indicated that 
the side-attached Rashba chain leads to competition between 
the Kondo and Majorana states \cite{Vernek-2015,Lee-2013,Cheng-2014,
Beek-2016,Weymann-2017,Wojcik-2017}. For sufficiently long wire 
($\epsilon_m=0$) the Kondo effect survives only in the spin-channel 
$\downarrow$, whereas for $\uparrow$ electrons there appears a dip 
in the spectral density at $\omega=0$. The resulting tunneling 
conductance is then partly reduced (from the perfect value $2e^2/h$) 
to the fractional value $3e^2/2h$ 
\cite{Vernek-2015,Lee-2013,Weymann-2017,Wojcik-2017,Lopez-2014}.
On contrary, for the short Rashba wires (with $\epsilon_m\neq0$) 
the Kondo physics survives in both spin channels. 

In our present setup (Fig.\ \ref{schematics}) the correlated quantum dot 
is between the metallic and superconducting reservoirs, therefore 
the Kondo effect is additionally affected by the on-dot pairing. 
Its influence is mainly controlled by the ratio $U/\Gamma_{\rm S}$ 
and partly by the level $\epsilon$, deciding whether the QD 
ground-state is in the spinful or spinless  configuration 
\cite{Domanski-2016,Zitko-2015a,Yamada-2011,Tanaka-2007,Baranski-2013}. 
Obviously the latter one cannot be screened. For instance, for the 
half-filled QD ($\epsilon=-\frac{U}{2}$) the spinful (doublet) 
configuration occurs in the regime $U \geq \Gamma_{\rm S}$.

\begin{figure}  
\includegraphics[width=0.8\linewidth]{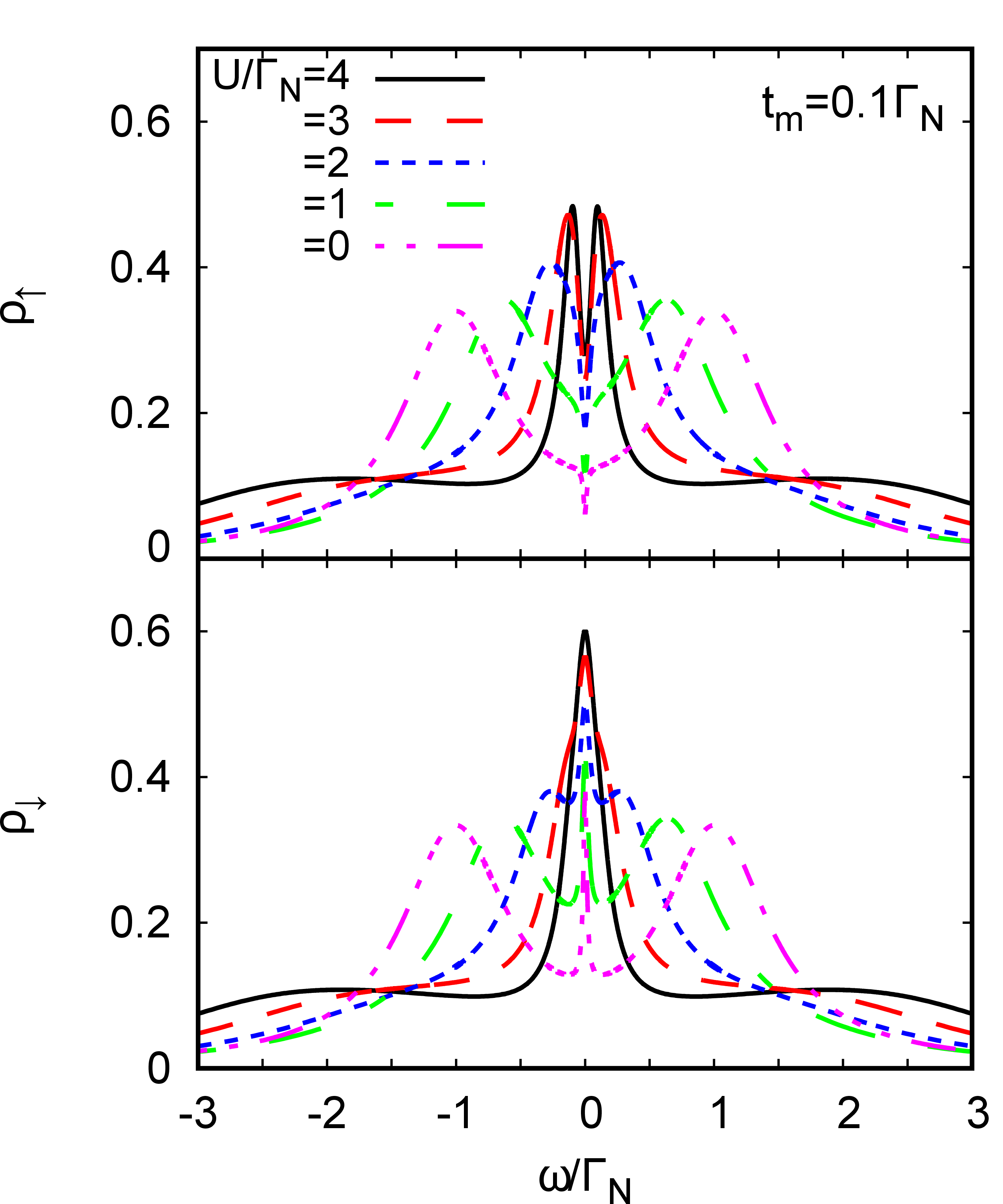}
\caption{The polarized spectral function $\rho_{\sigma}(\omega)$ 
obtained at zero temperature for the half-filled QD 
($\varepsilon=-U/2$), $\Gamma_S=2\Gamma_N$, $t_m=0.1\Gamma_{\rm N}$ 
and several values of the Coulomb potential $U$ (as indicated).
Energies are expressed in units of $\Gamma_{\rm N}$.}
\label{Kondo_spectrum_bothspins}
\end{figure}

For studying the correlations we adopt perturbative treatment 
of the Coulomb potential, treating it selfconsistently to
the second order in the normal and anomalous channels
 \cite{Vecino-2003,Yamada-2011}. Specific expressions 
have been given by us in Ref.\ \cite{Domanski-2016}.
Fig.\ \ref{Kondo_spectrum_bothspins} shows the spectral function 
$\rho_{\sigma}(\omega)$ for both spins obtained at zero 
temperature for the Coulomb potential $U$, covering the (spinless) 
singlet and (spinful) doublet configurations. In the weak interaction 
regime we observe appearance of two YSR
states. For 
$U\approx \Gamma_{\rm S}$ these peaks merge, signaling the quantum phase 
transition. The Kondo effect shows up only in the correlated limit 
($U>\Gamma_{\rm S}$), but its spectroscopic signatures are qualitatively 
different for each of the spins. Leakage of the Majorana quasiparticle
suppresses the low energy states of $\uparrow$ electrons. We notice
that the initial density (for $t_{m}=0$) is reduced by half, whereas
we observe a constructive influence of the Majorana quasiparticle on 
opposite spin $\downarrow$ electrons.

\begin{figure} 
\includegraphics[width=0.8\linewidth]{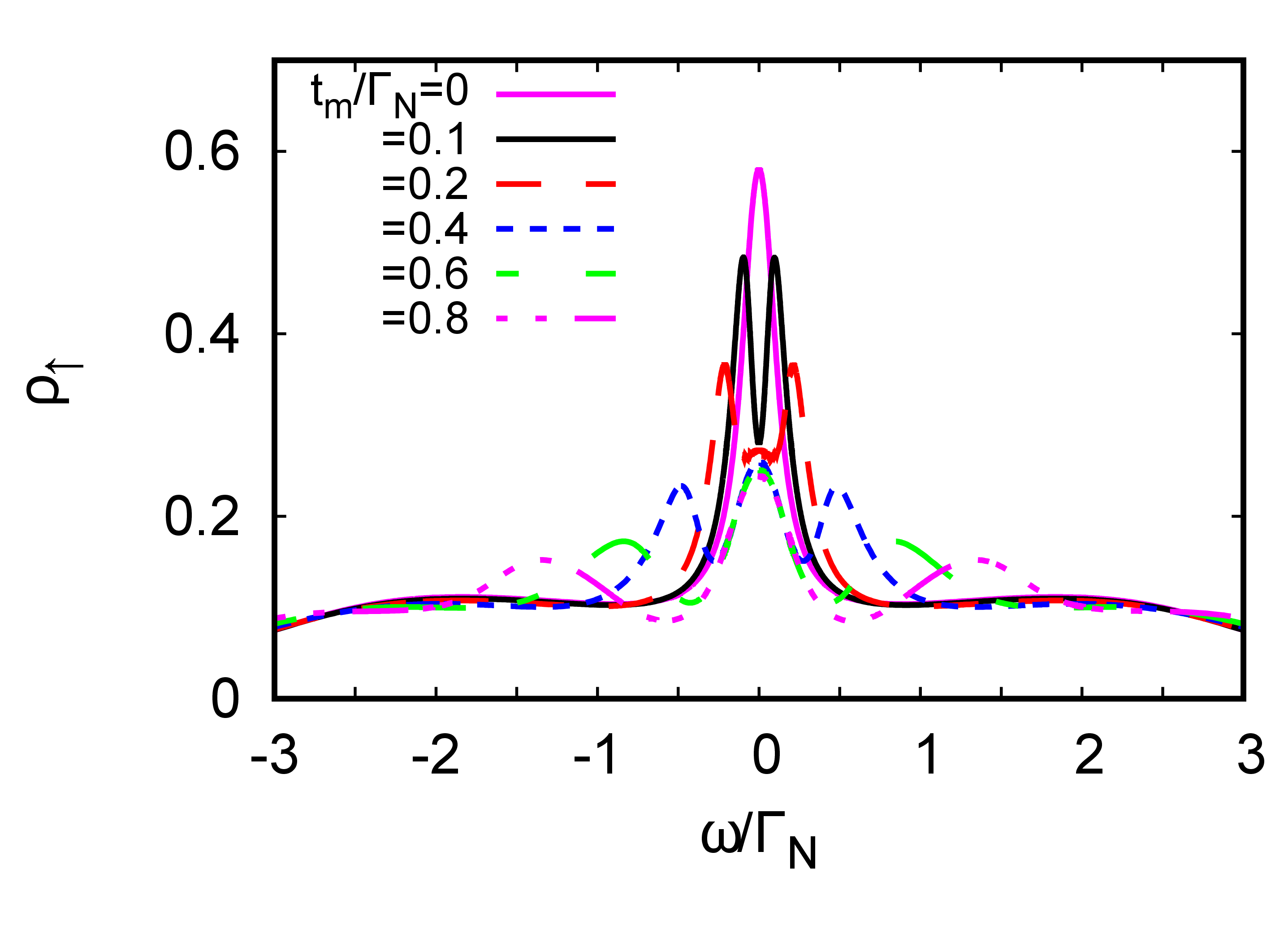}
\caption{The spectral function $\rho_{\uparrow}(\omega)$ of 
the half-filled quantum dot ($\varepsilon=-U/2$) obtained 
at $T=0$ for $\Gamma_S/\Gamma_{\rm N}=2$, $U/\Gamma_{\rm N}=4$ and several
values of  $t_m$ (as indicated). }
\label{Kondo_tm}
\end{figure}

Fig.\ \ref{Kondo_tm} shows evolution of the spectral function
$\rho_{\uparrow}(\omega)$ for various couplings $t_{m}$.
In the weak coupling limit we clearly observe reduction (by half) 
of the initial density of states. With increasing $t_{m}$ the spectrum
develops the three-peak structure, typical for the  'molecular' limit.
Such behavior indicates that the Majorana and Kondo states have 
rather a complicated 
relation, which is neither competitive
nor cooperative. In fact, some novel scaling laws have been
recently reported by several authors 
\cite{Beri-2012,Cheng-2014,Logan-2014,Beek-2016,Tsvelik-2016,Beri-2017}
although considering the correlation effects directly in the Rashba nanowire.

\section{Conclusions}
\label{sec.sum}

We have studied the polarized bound states of magnetic impurities embedded in the $s$-wave superconducting material, taking into account the spin-orbit and/or Coulomb  interactions.
We have shown that SO coupling strongly affects the subgap states,
both of the single impurities and their conglomerates arranged into 
nanoscopic chain. For the case of single magnetic impurity the
SO interaction (i) shifts the quantum phase transition towards higher magnetic coupling $J_{c}$, (ii) enhances spatial size of the YSR states, and (iii) smoothens the 
particle-hole oscillations. For the magnetic
chain such SO coupling combined with the Zeeman term induce the topologically nontrivial superconducting state and indirectly give rise to substantial polarization of the Majorana
modes (Fig. \ref{Rashba}), whose oscillations show up near the chain edges 
(Fig. \ref{oscillations}). The polarized Majorana quasiparticles can also leak onto other side-coupled objects, like the single or
multiple quantum impurities (Fig. \ref{proxy}). These polarized Majorana quasiparticles can be controlled 
by the magnetic field or by electrostatic potential (that would be important for future quantum computations using the qubits based on topologically protected Majorana states). Finally, we have also confronted the 
Majorana quasiparticles with the subgap Kondo effect, revealing
their complex relationship which can be hardly regarded as competitive or collaborative in some analogy to the Kondo effect originating from
multiple degrees of freedom \cite{Palacios-2013}. The aforementioned spin-polarized effects can be experimentally verified by the polarized ballistic tunneling or using STM spectroscopy, relying on the selective equal spin Andreev reflections.

\begin{acknowledgments}
We thank for instructive remarks from R.\ Aguado, J. Klinovaja, 
R.\ Lutchyn, P.\ Simon, and R.\ \v{Z}itko on different parts of
our study.
This work was supported by the National Science Centre (Poland) under grants 
DEC-2014/13/B/ST3/04451 (AK, SG, TD) and DEC-2013/11/B/ST3/00824 (MMM)
UMO-2017/25/B/ST3/02586 (AP)
and by
the Faculty of Mathematics and Natural Sciences
of the University of Rzesz\'ow through the project 
WMP/GD-06/2017 (GG).
\end{acknowledgments}

\bibliography{biblio}

\end{document}